\documentclass[aps,prd,amsmath,showpacs,nofootinbib]{revtex4}
\RequirePackage[]{amssymb}
\usepackage{graphicx,euscript}

\def\({\left(}
\def\){\right)}

\newcommand{\beq}{\begin{equation}}
\newcommand{\eeq}{\end{equation}}
\newcommand{\bea}{\begin{eqnarray}}
\newcommand{\eea}{\end{eqnarray}}

\newcommand{\bean}{\begin{eqnarray*}}
\newcommand{\eean}{\end{eqnarray*}}
\newcommand{\bs}{\begin{subequations}}
\newcommand{\es}{\end{subequations}}

\newcommand{\diag}{\textrm{diag}}

\begin{document}

\title{Non-vanishing cosmological constant effect \\
in super-Poincare-invariant  Universe.}

\author{Asya V. Aminova}
\email{asya.aminova@kpfu.ru}
\affiliation{Department of General Relativity and Gravitation, 
 Kazan
Federal University,  18 Kremlyovskaya St., Kazan 420008,
Russia}

\author{Mikhail Kh. Lyulinsky}%
\email{miklul@rambler.ru}%
\affiliation{Department of General Relativity and Gravitation, 
 Kazan
Federal University,  18 Kremlyovskaya St., Kazan 420008,
Russia}

\begin{abstract}

In  \cite{AminMoc} we defined the Minkowski superspace $SM(4,4\vert \lambda, \mu)$    as the  invariant of the Poincare
supergroup of  supertransformations, which is a solution of Killing superequations.
   In the present
    paper we use  formulae of super-Riemannian geometry developed by V.~P. Akulov and D.~V. Volkov  \cite{AkVolk} for  calculating a superconnection and a supercurvature of  Minkowski superspace. We  show
    that the curvature of the Minkowski superspace does not vanish,  and the  Minkowski supermetric is the solution of the Einstein superequations,
so the eight-dimensional curved  super-Poincare invariant superuniverse $SM(4,4\vert \lambda, \mu)$   is supported
by purely fermionic stress-energy supertensor with two real parameters $\lambda$, $\mu$, and, moreover, it has non-vanishing cosmological constant $\Lambda=12/(\lambda^2 -\mu^2)$ defined by these parameters that could mean a new look at the cosmological constant problem.

\end{abstract}


 \maketitle

 \maketitle

\section{Introduction.}

The consistent supersymmetry approach to the  theory of gravitation
according to which supergeometry must be determined  by the properties of supersymmetry requires the developing of group-invariant  methods  in the
supergravity. In this direction not only there were absent   concrete  results
 but in many cases the very concepts that must
be  basic for the connection between the supergeometry and
supersymmetry have not been developed.
In  \cite{AminMoc} an attempt was made to fill in this gap.

We consider the supersymmetry as an  automorphism of a supergeometric structure and, in part, as an infinitesimal
supertransformation preserving the metric  of the  superspace, the metric itself is defined as the  invariant of the corresponding supergroup of transformations, in the spirit of Klein's approach where the notion of  symmetry, or group of transformations,  is  a  fundamental  notion determining  the geometry of the space. Following this program
we derived  in \cite{AminMoc} the Lie superderivative of a supermetric and defined the Minkowski superspace $SM(4,4\vert \lambda, \mu)$  as an  invariant of the Poincare
supergroup of  supertransformations, which is a solution of Killing superequations. The found supermetric contains two real parameters $\lambda$, $\mu$ and becomes degenerate only if $\lambda=\pm \mu$. As a result we obtained the two-parameter family of the Minkowski  superuniverses that is being studied in this article.

The article is organised as follows. In the first section we introduce basic
definitions and give a brief review of earlier results.
In the second section we
discuss the super-Poincare-invariant Minkowski superspace.
The third and fourth sections are  devoted to calculating a superconnection  and a supercurvature of the Minkowski superspace by using  the apparatus of super-Riemannian geometry developed by V.~P. Akulov and D.~V. Volkov's  \cite{AkVolk}. In the fifth section we write  and analyze  the Einstein superequations for Minkowski superspace.

We recall that a linear space $g$ is called $\mathbb{Z}_2$-graded if it is represented in the form of a direct sum of two spaces: $g=g^0\oplus g^1$; elements of $g^0$ and $g^1$ are called homogeneous elements, even and odd, respectively. The dimension of $g$ is a pair $(p,q)$, where $p$ is dimension of even subspace and $q$ is dimension of odd subspace. The fact that $x\in g^i$, \ $i\in \mathbb{Z}_2$ \ $(i=0,1)$ is written in the form $\sigma(x)=i$ and $\sigma(x)$ is called the \textit{parity} of the element $x$.

The Lie superalgebra is a $\mathbb{Z}_2$-graded linear space with a fixed parity
$g=g^0\oplus g^1$, on which a bilinear operation $[x,y]$ (the supercommutator) is given so that for any homogeneous elements $x,y,z \in g$ there hold the following identities:
\begin{equation*}
\sigma([x,y])=\sigma(x)+\sigma(y),
\end{equation*}
\begin{equation*}
[x,y]=(-1)^{\sigma(x)\sigma(y)+1}[y,x],
\end{equation*}
\begin{equation*}
[x,[y,z]](-1)^{\sigma(x)\sigma(z)}+[z,[x,y]](-1)^{\sigma(z)\sigma(y)}+[y,[z,x]](-1)^{\sigma(y)\sigma(x)}=0,
\end{equation*}

Let $\{e_A\}$ be a standard homogeneous basis in $g$: $\{e_A\}\in g^0$ for $1\leq A\leq p$ and $\{e_A\}\in g^1$ for  $p+1\leq A \leq p+q$, further we use the notation
\begin{equation*}
\sigma(A)=\sigma(e_A).
\end{equation*}
 The structure constants $C_{AB}^D$ of the Lie superalgebra are defined by the expansion
$[e_A,e_B]=C_{AB}^De_D$.

Let $\Lambda_q$ be the Grassmannian algebra, i.e. an associative algebra with the unit, where there exists a system of $q$ generators $\theta^1,\ldots,\theta^q$ satisfying the relations
\begin{equation}\label{grass}
\theta^\alpha\theta^\beta+\theta^\beta\theta^\alpha=0, \ \left(\theta^\alpha\right)^2=0 \quad (\alpha,\beta =1,\ldots,q).
\end{equation}
 If a multiplication of elements of the Lie superalgebra $g$ by elements of $\Lambda_q$ from the left is defined and for homogeneous elements $\mu,\nu\in \Lambda_q$ and $x,y\in g$ there holds
\begin{equation*}
[\mu x,\nu y]=\mu\nu[x,y](-1)^{\sigma(\nu)\sigma(x)},
\end{equation*}
then $g$ is called the Lie superalgebra with Grassmannian structure.

The Lie superalgebras and there  Grassmannian spans can be  realized as an algebra of differential operators of the form
\begin{equation}\label{oper}
(Xf)(z)=\sum_A X^A(z)\frac{\partial}{\partial z^A}f(z),
\end{equation}
where $X^A(z)$ and $f(z)$ belong to the local supermanifold, i.e. to the algebra $\Lambda_{p,q}(U)$ of functions defined on $U\subset \mathbb{R}^p$ with their values in Grassmannian algebra $\Lambda_q$. Elements of the algebra $\Lambda_{p,q}(U)$  can be written in the form
\begin{equation}\label{Lam_pq}
f=f(x,\theta)=\sum_{k\geq 0}\sum _{\alpha_1,\ldots,\alpha_k} f_{\alpha_1\ldots   \alpha_k}(x^1,\dots,x^p)\theta^{\alpha_1}\ldots\theta^{\alpha_k},
\end{equation}
where $x^1,\ldots,x^p$ are coordinates in $U$, $\theta^{1},\ldots,\theta^{q}$ are generators of $\Lambda_q$, and $(z^1,\ldots,z^{p+q})\equiv (x^1,\ldots,x^p,\theta^{1},\ldots,\theta^{q})$ are the homogeneous (i.e. even $(x)$ and odd $(\theta)$) generators of  $\Lambda_{p,q}(U)$. In the case of odd generator $z^A=\theta^A$, i.e. when $\sigma (z^A)=1$, the symbol $\partial/\partial z^A$ in \eqref{oper} denotes the left derivative which is computed by carrying $\theta^A$ out from the product $\theta^\alpha\ldots\theta^\beta$  to the left according to rules \eqref{grass} and then deleting.

The set  $D_{p,q}(U)$ of operators \eqref{oper} is $\mathbb{Z}_2$-graded. An operator $X$ is homogeneous if its coefficients $X^A$ are homogeneous, and the sum $\sigma(X^A)+\sigma(z^A)$ does not depend on $A$. In this case, the parity of the operator $X$ is equal to $\sigma(X)=\sigma(X^A)+\sigma(z^A)$.

The supercommutator (the super Lie bracket)
\begin{equation*}
[ X,Y]=XY-(-1)^{\sigma(X)\sigma(Y)}YX,
\end{equation*}
which defines a superanalog of the Lie derivative $L_X Y$ turns  $D_{p,q}(U)$ into the Lie superalgeba called the Lie superalgebra of (local) vector fields. For homogeneous operators $X,Y\in D_{p,q}(U)$ we have
\begin{equation*}
[X,Y]^A=\sum_C\left(X^C \frac{\partial}{\partial z^C}Y^A-(-1)^{\sigma(X)\sigma(Y)}
Y^C \frac{\partial}{\partial z^C}X^A\right).
\end{equation*}

\section{The super-Poincare-invariant Minkowski superspace.}

Let $M_0$ be a $p$-dimensional differentiable manifold. We assign to each open subset $U\subset M_0$ the algebra $\Lambda_{p,q}(U)$ of infinitely diferentiable functions (\ref{Lam_pq}) on $U$ with values in Grassmannian algebra $\Lambda_q$. The manifold $M_0$ with the sheaf of algebras $\Lambda_{p,q}(U)$ is a supermanifold.

Consider a supermanifold with coordinates $z^{A}=(x^{a}, \theta^{\alpha })$, where $x^{a}$ $(a=0,1,2,3)$ are bosonic (even), and $\theta^{\alpha }$ $(\alpha=1,2,3,4)$ are  fermionic (odd) coordinates. Then
$\sigma(x^a)=0$, $\sigma(\theta^\alpha)=1$, where $\sigma$ is the parity operator.

A superspace is said to be  Riemannian one or super-Riemannian space if on this superspace there is given a non-degenerate metric form (\cite{Berezin}, p. 22, and \cite{AkVolk})
\begin{equation}\label{eq1}
ds^2=dz^{B} dz^{A} g_{AB}(z)\equiv \tilde g_{AB}dz^Adz^B, \quad g_{AB}(z)=(-1)^{\sigma(A) \sigma(B)} g_{BA}(z),
\end{equation}
where differentials $dz^A$ and $dz^B$ possess the same parity as the corresponding coordinates do, and
\begin{equation*}
\tilde g_{AB}= (-1)^{\sigma(A)+\sigma(B)+\sigma(A)\sigma(B)}g_{AB}.
\end{equation*}

The contravariant metric tensor $g^{AB}$ is determined by the relation
\begin{equation}\label{contr}
g^{AC} g_{CB}=(-1)^{\sigma(A) \sigma(B)} {\delta_{B}}^{A}.
\end{equation}
Thanks to the signature factor $(-1)^{\sigma(A) \sigma(B)}$ in (\ref{contr}) we have
\begin{equation}\label{sum}
g^{AC} g_{CA}=(-1)^{\sigma(A)} {\delta_{A}}^{A}=0.
\end{equation}

Minkowski superspace was defined in \cite{AminMoc} as a superspace  endowed with a supermetric invariant with respect to transformations belonging to the Poincare's supergroup. Then the Minkowski suppermetric must satisfy super Killing's equations
\begin{equation*}
\begin{array}{c}
L_Xg_{AB}\equiv Xg_{AB}+(-1)^{\sigma(A)\sigma(X)}\left(\partial_AX^C\right)g_{CB}+
(-1)^{\sigma(B)[\sigma(A)+\sigma(X)]}\left(\partial_BX^C\right)g_{CA}=0,
\end{array}\end{equation*}
where $L_X{}g_{AB}$  is the Lie derivative $L_X$ of a metric form $g_{AB}$ with respect to $X$ \cite{AminMoc}, $X$ are generators of the Poincare's supergroup realized in the following infinitesimal transformations:
\begin{equation*}
P_a=\partial_a, \quad  M_{ab}=x_a\partial_b-x_b\partial_a + \frac{1}{2}\theta^\alpha\left(\gamma_{ab}\right)_\alpha{}^\beta\partial_\beta, \quad Q_\alpha =\partial_\alpha-\theta^\beta\left(\gamma^aC^{-1}\right)_{\beta\alpha}\partial_a,
\end{equation*}
 hereinafter lowercase Latin indices $a,b,\ldots$ take the values $0,1,2,3$, lowercase Greek indices $\alpha,\beta,\ldots =1,2,3,4$, $\gamma^a$ are the Dirac gamma-matrices, $\gamma^{ab}=(1/2)(\gamma^a\gamma^b-\gamma^b\gamma^a),$ and $C=i\gamma^0\gamma^2$ is the matrix of charge conjugation.

Solving the super Killing equations with respect to $g_{AB}$, we have obtained the superextension of the Minkowski metric $\eta_{ab}=\diag(1,-1,-1,-1)$ as a supermetric invariant under the Poincare's supergroup
\begin{equation}\label{SM}\begin{array}{c}
ds^{2}_{SM}=dx^{b}dx^{a}\eta_{ab}-2d\theta^{\beta}dx^{a}\theta^{\alpha}\left(\gamma^{a}C^{-1}\right)_{\beta\alpha}
+
\\ \\
d\theta^{\beta}d\theta^{\alpha}\left(\theta^{\gamma}\theta^{\delta}\left(\gamma^{a}C^{-1}
\right)_{\alpha\gamma}
\left(\gamma^{a}C^{-1}\right)_{\beta\delta}+{\lambda}\left(C^{-1}\right)_{\alpha\beta}
+{\mu}\left(\gamma_{5}C^{-1}\right)_{\alpha\beta}\right). \end{array}\end{equation}
For finding contravariant components of Minkowski supermetric we expand   the matrix $\left(g_{AB}\right)$ in a series with respect to variables $\theta$ and separate the addend of degree zero: $g=\stackrel{0}{g}+U$, where
\begin{equation}
\label{eq19}
\stackrel{0}{g}=\left(\begin{array}{cc}
\eta_{ab} & \ \ 0\\ \\
0 & \ \  \lambda\left(C^{-1}\right)_{\alpha\beta}+\mu\left(\gamma_{5}C^{-1}\right)_{\alpha\beta}
\end{array}\right),
\end{equation}
\begin{equation*}
U=\left(\begin{array}{cc}
0 & \ \ -\theta^{\delta}\left(\gamma_{b}C^{-1}\right)_{\alpha\delta}\\ \\
-\theta^{\delta}\left(\gamma_{a}C^{-1}\right)_{\beta\delta} &  \ \ \theta^{\gamma}\theta^{\delta}\left(\gamma^{a}C^{-1}\right)_{\alpha\gamma}\left(\gamma_{a}C^{-1}\right)
_{\beta\delta}
\end{array}\right).
\end{equation*}
The inverse to $g$ matrix $g^{-1}=\left(g^{AB}\right)$  exists if
and only  if there exists an inverse to $\stackrel{0}{g}$ matrix $\stackrel{0}{g}{}^{-1}$ (\cite{Berezin}, p.~91). In this case the inverse matrix $g^{-1}$  can be found
with the help of the formula
\begin{equation}
\label{eq21}
g^{-1}=\left(\sum_{n=0}^{\infty}\left(-1\right)^{n}\left(\stackrel{0}{g}{}^{-1}U\right)^{n}\right)
\stackrel{0}{g}{}^{-1}
\end{equation}
where the series from the right is finite because of nilpotency of
the matrix $\left(\stackrel{0}{g}{}^{-1}U\right)$.
It is easy to check that the inverse of the block-diagonal matrix (\ref{eq19}) will not exist
only in the following two cases:
$
a)\ \lambda=\mu,
\quad
b)\  \lambda=-\mu.$
 In all other cases the inverse of the matrix  $\stackrel{0}{g}$ is of the form
\begin{equation*}
\stackrel{0}{g}{}^{-1}=\left(\begin{array}{cc}
\eta^{ab} & 0\\ \\
0 &  \left(\lambda^{2}-\mu^{2}\right)^{-1}\left(\lambda C^{\alpha\beta}-\mu\left(C\gamma_{5}\right)^{\alpha\beta}\right)
\end{array}\right).
\end{equation*}
Applying  (\ref{eq21}) and taking into account (\ref{contr}) and (\ref{sum})
 we find that
\begin{equation*}
g^{-1}=\left(g^{AB}\right)=\left(\begin{array}{cc}
g^{ab} & g^{a\beta}\\
g^{\alpha b} & g^{\alpha\beta}
\end{array}\right),\quad
\end{equation*}
where
\begin{equation}\label{ContrComp}
g^{ab}=\eta^{ab}\left(1+\theta^{\mu}\theta^{\nu}A_{\mu\nu}\right), \quad g^{a\beta}=\theta^{\mu}\left(L^{a}\right)\mu^{\beta},\quad
g^{b \alpha} = \theta^{\mu}({L^b}{}^T)^{\alpha}{}_{\mu}, \quad
g^{\alpha \beta} = - B^{\alpha \beta},
\end{equation}
and the following notations are used:
\begin{equation}\label{ContrCompA}
A_{\alpha \beta} \equiv \frac{1}{\lambda^2 - \mu^2}\left(\lambda(C^{-1})_{\alpha \beta} + \mu (\gamma_5 C^{-1})_{\alpha \beta}\right),
\end{equation}

\begin{equation}\label{ContrCompL}
(L^b)_{\alpha}{}^{ \beta} = \frac{1}{\lambda^2 - \mu^2}\left(\lambda (\gamma^{b})_{\alpha}{}^{ \beta} - \mu(\gamma_5 \gamma^b )_{\alpha}{}^{ \beta}\right),
\end{equation}

\begin{equation}\label{ContrCompB}
B^{\alpha \beta} = \frac{1}{\lambda^2 - \mu^2} \left(\lambda C^{\alpha \beta} - \mu (C\gamma_5)^{\alpha \beta}\right).
\end{equation}
If following the authors of the book \cite{Green} we assume that the "real" theory of a massless point particle is a supersymmetric theory \cite{Green} (Vol. 1, p. 33) then we must replace the action of a classical massless point particle
\begin{equation*}
S=\int\eta_{ab}dx^adx^b
\end{equation*} by its supersymmetric generalization
\begin{equation}\label{SMact}
S=\int ds^2_{SM}
\end{equation}
where $ds^2_{SM}$ is the supermetric (\ref{SM}) derived  from group-theoretic considerations. It is important to note that, given the  rule $\int d\theta^\alpha=0$ (\cite{Berezin}, p.~74) the action (\ref{SMact})
corresponds to the action
\begin{equation}\label{GSWact}
S=\int \eta_{ab}\left(dx^a+\theta^\alpha\left(C\gamma^a\right)_{\beta\alpha}d\theta^\beta\right)
\left(dx^b+\theta^\mu\left(C\gamma^b\right)_{\nu \mu}d\theta^\nu\right)
\end{equation}
(\cite{Green}, the formulae (1.3.4), p.~32 and  (5.1.5), p.~283)
 that describes a point particle propagating not in Minkowski space, but in a superspace with  coordinates $(x^a,\theta^\alpha)$, here $\theta^\alpha$ are anticommuting coordinates, transforming as spinors under Lorentz transformations of coordinates $x^a$.

\section{A superconnection of the Minkowski superspace.}

V.~P. Akulov and D.~V. Volkov \cite{AkVolk} have defined   the Levi-Civita superconnection
of the super-Riemannian metric (\ref{eq1}) by the formulae
\begin{equation*}
{\Gamma_{A}}^{B}(d)=dz^{C} {\Gamma_{CA}}^{B}(z),
\end{equation*}
\begin{equation*}
\Gamma_{AB}(d)={\Gamma_{A}}^{C}(d) g_{CB}=dz^{D} \Gamma_{D,AB},
\end{equation*}
\begin{equation*}
\Gamma_{A,BC}=\frac{1}{2} \left( \partial_{A} g_{BC}+ (-1)^{\sigma(A) \sigma(B)} \partial_{B} g_{AC} - (-1)^{\left( \sigma(A) + \sigma(B)\right) \sigma(C)} \partial_{C} g_{AB}\right),
\end{equation*}
\begin{equation}
\label{eq5}
{\Gamma_{A,B}}^{C}=(-1)^{\sigma(D)} \Gamma_{A,BD} g^{DC},
\end{equation}
where the following  symmetry properties  hold:
\begin{equation*}
\Gamma_{A,BC}=(-1)^{\sigma(A) \sigma(B)} \Gamma_{B,AC}.
\end{equation*}

Applying these formulae to Minkowski supermetric (\ref{SM}) we get
\begin{equation*}
\Gamma_{a,bc} = \Gamma_{a, b \delta} = \Gamma_{a, \beta c} = \Gamma_{\alpha, b c} = \Gamma_{\alpha, \beta c} = 0,
\end{equation*}

\begin{equation*}
\Gamma_{a, \beta \delta} = \Gamma_{\beta, a \delta} = -(\gamma_a C^{-1})_{\beta \delta},
\end{equation*}

\begin{equation*}
\Gamma_{\alpha, \beta \delta} = \theta^{\nu}\left((\gamma_a C^{-1})_{\nu \alpha}(\gamma^a C^{-1})_{\beta \delta} - (\gamma_a C^{-1})_{\nu \beta}(\gamma^a C^{-1})_{\alpha \delta}\right).
\end{equation*}
For $\Gamma_{A,B}{}^C$ we find by using (\ref{eq5}) and (\ref{ContrComp})--(\ref{ContrCompB})
\begin{equation*}
\Gamma_{a,b}{}^c = \Gamma_{a, b}{}^{\delta} = 0,
\end{equation*}

\begin{equation*}
\Gamma_{a, \beta}{}^c = \Gamma_{\beta, a}{}^c = -\theta^{\mu}(M_a{}^c)_{\beta \mu},
\end{equation*}

\begin{equation*}
\Gamma_{a, \beta}{}^{\delta} = \Gamma_{\beta, a}{}^{\delta} = -(L_a)_{\beta}{}^{\delta},
\end{equation*}

\begin{equation*}
\Gamma_{\alpha, \beta}{}^c = \theta^{\mu} \theta^{\nu}\left((\gamma^a C^{-1})_{\mu \alpha} (M_a{}^c)_{\beta \nu} - (\gamma^a C^{-1})_{\mu \beta} (M_a{}^c)_{\alpha \nu}\right),
\end{equation*}

\begin{equation*}
\Gamma_{\alpha, \beta}{}^{\delta} = \theta^{\nu}\left((\gamma^a C^{-1})_{\nu \alpha} (L_a)_{\beta}{}^{\delta} - (\gamma^a C^{-1})_{\nu \beta} (L_a)_{\alpha}{}^{\delta}\right),
\end{equation*}

where
$$(M_a{}^c)_{\alpha \beta} = \frac{1}{\lambda^2 - \mu^2}\left(\lambda (\gamma_a\gamma^c C^{-1})_{\alpha \beta} + \mu (\gamma_a \gamma_5 \gamma^c C^{-1})_{\alpha \beta}\right) .$$

\section{A supercurvature of the Minkowski superspace.}

A curvature supertensor of the Riemannian supermetric (\ref{eq1}) is defined by the formula \cite{AkVolk}
\begin{equation*}\begin{array}{c}
{R_{DBA}}^{C}=(-1)^{\sigma(D) \sigma(B)} \partial_{B} {\Gamma_{DA}}^{C}- \partial_{D} {\Gamma_{BA}}^{C}+(-1)^{\sigma(A) \sigma(B)}  {\Gamma_{DA}}^{F} {\Gamma_{FB}}^{C} - \\ \\
(-1)^{\left( \sigma(A) + \sigma(B)\right) \sigma(D)}  {\Gamma_{BA}}^{F} {\Gamma_{FD}}^{C}
\end{array}\end{equation*}
and has the  symmetry properties
\begin{equation*}
R_{DCBA}=-(-1)^{\sigma(A) \sigma(B)}R_{DCAB}=-(-1)^{\sigma(C) \sigma(D)}R_{CDBA},
\end{equation*}
\begin{equation*}
R_{DCBA}=(-1)^{\left( \sigma(A) + \sigma(B)\right)\left( \sigma(C) + \sigma(D)\right)} R_{BADC},
\end{equation*}
\begin{equation*}
R_{DCBA}+(-1)^{\sigma(B)\left( \sigma(C) + \sigma(D)\right)} R_{BDCA}+(-1)^{\left( \sigma(C) + \sigma(B)\right) \sigma(D)} R_{CBDA}=0,
\end{equation*}
where
\begin{equation*}
R_{DCBA}={R_{DCB}}^{F} g_{FA}.
\end{equation*}

Using these formulae and the superconnection ${\Gamma_{A,B}}^{C}$ of Minkowski supermetric (\ref{SM}) found in the preceding section we have
\begin{equation*}
R_{dba}{}^c = R_{dba}{}^{\tau} = 0,
\end{equation*}
\begin{eqnarray*}
&R_{db\alpha}{}^c=\dfrac{2\theta^{\nu}}{\lambda^2-\mu^2}\left( \gamma_{db}\gamma^{c}C^{-1} \right)_{\alpha\nu},&\\
&R_{d\beta a}{}^c=-R_{\beta d a}{}^c=-\dfrac{\theta^{\nu}}{\lambda^2-\mu^2}\left( \gamma_{a}\gamma_{d}\gamma^{c}C^{-1} \right)_{\beta\nu},&\\
&R_{db\alpha}{}^{\delta}=\dfrac{2}{\lambda^2-\mu^2}\left( \gamma_{db}\right)_{\alpha}{}^{\delta},&\\
&R_{d\beta a}{}^{\delta}=-R_{\beta d a}{}^{\delta}=-\dfrac{1}{\lambda^2-\mu^2}\left( \gamma_{a}\gamma_{d} \right)_{\beta}{}^{\delta},&
\end{eqnarray*}

\begin{eqnarray*}
&R_{\delta b\alpha}{}^{c}=-R_{b\delta\alpha}{}^{c}=&\\ \\
&\left(M_b{}^c\right)_{\alpha\delta}+
\dfrac{\theta^{\varepsilon}\theta^{\nu}}{\lambda^2-\mu^2}
\left(2\left(\gamma^k C^{-1}\right)_{\varepsilon\delta}\left(\gamma_{bk}\gamma^cC^{-1}\right)_{\alpha\nu}+\right.&\\ \\
&\left.\left(\gamma^kC^{-1}\right)_{\varepsilon\alpha}\left(\gamma_k \gamma_b \gamma^c C^{-1}\right)_{\delta\nu}\right),&
\end{eqnarray*}

\begin{eqnarray*}
&R_{\delta \beta a}{}^{c}=\left(M_a{}^c\right)_{\delta\beta}+\left(M_a{}^c\right)_{\beta\delta}+&\\ \\ &
\dfrac{\theta^{\varepsilon}\theta^{\nu}}{\lambda^2-\mu^2}
\left(\left(\gamma^k C^{-1}\right)_{\varepsilon\beta}\left(\gamma_{a}\gamma_k\gamma^c C^{-1}\right)_{\delta\nu}+\right.
& \\ \\&
\left.\left(\gamma^kC^{-1}\right)_{\varepsilon\delta}\left(\gamma_a \gamma_k \gamma^c C^{-1}\right)_{\beta\nu}\right),&
\end{eqnarray*}

\begin{eqnarray*}
&R_{\delta \beta \alpha}{}^{c}=
\theta^{\varepsilon}\left(
-2\left(\gamma^k C^{-1}\right)_{\beta\delta}\left(M_k{}^c\right)_{\alpha\varepsilon}+
\left(\gamma^k C^{-1}\right)_{\beta\alpha}\left(M_k{}^c\right)_{\delta\varepsilon}+\right.
&\\ \\&
\left(\gamma^k C^{-1}\right)_{\delta\alpha}\left(M_k{}^c\right)_{\beta\varepsilon}+
\left(\gamma^k C^{-1}\right)_{\varepsilon\delta}\left(M_k{}^c\right)_{\alpha\beta}+
\left(\gamma^k C^{-1}\right)_{\varepsilon\beta}\left(M_k{}^c\right)_{\alpha\delta}-
&\\ \\&  \left.
-\left(\gamma^k C^{-1}\right)_{\varepsilon\alpha}\left(M_k{}^c\right)_{\delta\beta}
-\left(\gamma^k C^{-1}\right)_{\varepsilon\alpha}\left(M_k^{}c\right)_{\beta\delta}\right)+
&\\ \\&
\dfrac{\theta^{\varepsilon}\theta^{\nu}\theta^{\sigma}}{\lambda^2-\mu^2}
\left(
\left(\gamma^k C^{-1}\right)_{\varepsilon\delta} \left(\gamma^l C^{-1}\right)_{\nu\beta} \left(\gamma_k\gamma_l\gamma^c C^{-1}\right)_{\alpha\sigma}+\right.
& \\ \\&
\left(\gamma^k C^{-1}\right)_{\varepsilon\beta} \left(\gamma^l C^{-1}\right)_{\nu\delta} \left(\gamma_k\gamma_l\gamma^c C^{-1}\right)_{\alpha\sigma}-
& \\ \\&
\left(\gamma^k C^{-1}\right)_{\varepsilon\alpha} \left(\gamma^l C^{-1}\right)_{\nu\beta} \left(\gamma_k\gamma_l\gamma^c C^{-1}\right)_{\delta\sigma}-\nonumber
& \\ \\ &\left.
\left(\gamma^k C^{-1}\right)_{\varepsilon\alpha} \left(\gamma^l C^{-1}\right)_{\nu\delta} \left(\gamma_k\gamma_l\gamma^c C^{-1}\right)_{\beta\sigma}\right),&
\end{eqnarray*}

\begin{eqnarray*}
&R_{\delta \beta a}{}^{\tau}=\dfrac{\theta^{\nu}}{\lambda^2-\mu^2}
\left(
\left(\gamma^k C^{-1}\right)_{\nu\beta} \left(\gamma_a\gamma_k\right)_{\delta}{}^{\tau}+
\left(\gamma^k C^{-1}\right)_{\nu\delta} \left(\gamma_a\gamma_k\right)_{\beta}{}^{\tau}
\right),&
\end{eqnarray*}

\begin{eqnarray*}
&R_{\delta b \alpha}{}^{\tau}=-R_{b\delta\alpha}{}^{\tau}=\dfrac{\theta^{\nu}}{\lambda^2-\mu^2}\left(
2\left(\gamma^k C^{-1}\right)_{\nu\delta} \left(\gamma_{bk}\right)_{\alpha}{}^{\tau}+\right.
&\\ \\&\left.
\left(\gamma^k C^{-1}\right)_{\nu\alpha} \left(\gamma_k\gamma_b\right)_{\delta}{}^{\tau}\right),&
\end{eqnarray*}

$$
R_{\delta \beta \alpha}{}^{\tau} = -2{\left(\gamma^{c} C^{-1}\right)}_{\delta \beta} {{\left(L_c\right)}_{\alpha}}^{\tau} + {\left(\gamma^{c} C^{-1}\right)}_{\beta \alpha} {{\left(L_c\right)}_{\delta}}^{\tau} - {\left(\gamma^{c} C^{-1}\right)}_{\delta \alpha} {{\left(L_c\right)}_{\beta}}^{\tau}+$$ $$
 \frac{\theta^\epsilon \theta^\nu}{\lambda^2-\mu^2}\left({\left(\gamma^{c} C^{-1}\right)}_{\epsilon \delta} {\left(\gamma^{a} C^{-1}\right)}_{\nu \beta} {{\left(\gamma_{c} \gamma_{a}\right)}_\alpha}^{\tau} + {\left(\gamma^{c} C^{-1}\right)}_{\epsilon \beta} {\left(\gamma^{a} C^{-1}\right)}_{\nu \delta} {\left(\gamma_c \gamma_a\right)}_\alpha{}^{\tau} - \right.$$ $$ \left.  {\left(\gamma^{c} C^{-1}\right)}_{\epsilon \alpha} {\left(\gamma^{a} C^{-1}\right)}_{\nu \beta} {{\left(\gamma_{c} \gamma_{a}\right)}_\delta}^{\tau} -
  {\left(\gamma^{c} C^{-1}\right)}_{\epsilon \alpha} {\left(\gamma^{a} C^{-1}\right)}_{\nu \delta} {{\left(\gamma_{c} \gamma_{a}\right)}_\beta}^{\tau}\right).$$

\section{The  Einstein superequations.}

The Ricci tensor and the scalar curvature of the Riemannian supermetric (\ref{eq1})  are defined by the equations \cite{AkVolk}
\begin{equation*}
R_{BA}=(-1)^{\sigma(F)\left( \sigma(F) + \sigma(A)\right)} {R_{BFA}}^{F},
\end{equation*}
\begin{equation*}
R=g^{AB}R_{BA}.
\end{equation*}
From here using  the formulae of the section IV and the equations (\ref{ContrComp})--(\ref{ContrCompB}) we obtain
 Ricci curvature components of the  Minkowski supermetric (\ref{SM})
\begin{equation*}
R_{ab}=\frac{4}{\lambda^{2}-\mu^{2}} g_{ab},
\end{equation*}
\begin{equation*}
R_{b \alpha }=\frac{4}{\lambda^{2}-\mu^{2}} g_{b\alpha },
\end{equation*}
\begin{equation*}
R_{ \beta a}=\frac{4}{\lambda^{2}-\mu^{2}} g_{ \beta a},
\end{equation*}
\begin{equation*}
R_{\alpha \beta}=\frac{4}{\lambda^{2}-\mu^{2}} g_{\alpha \beta} + 4 A_{\alpha \beta}
\end{equation*}
(where $A_{\alpha \beta}$ is given by the equation (\ref{ContrCompA})) and finally a scalar curvature
\begin{equation*}
R=-\frac{16}{\lambda^{2}-\mu^{2}}.
\end{equation*}
By writing the Einstein equations with cosmological constant
\begin{equation}
\label{SEE}
R_{AB}-\frac{1}{2}R g_{AB}+\Lambda g_{AB}=T_{AB}
\end{equation}
we find that the Minkowski supermetric (\ref{SM}) is the solution of these equations with cosmological constant
$$
\Lambda=-\frac{12}{\lambda^{2}-\mu^{2}}
$$
and stress-energy supertensor
\begin{equation}\label{SET}
T_{AB}=\left(\begin{array}{cc}
0 & 0\\ \\
0 &  4(\lambda^2 - \mu^2)^{-1}\left(\lambda(C^{-1})_{\alpha \beta} + \mu (\gamma_5 C^{-1})_{\alpha \beta}\right)
\end{array}\right)
\end{equation}
with zero bosonic and nonzero fermionic parts. It is important to note that cosmological constant
is nonzero for any values of real parameters $\lambda, \mu$, $\lambda\neq \pm \mu$   of the Minkowski superspace-time family.

\section{Conclusion}
We follow earlier proposed in \cite{AminMoc}  scheme of obtaining  supergeneralizations of classical solutions of General Relativity. On the first stage, this scheme  includes constructing supersymmetric expansions of corresponding space-time symmetry groups, then finding solutions of supersymmetric Killing  equations and analyzing   corresponding Einstein superequations. In this paper the scheme is realized for the  Minkowski space whose symmetry group is the Poincare group and its supersymmetric expansion is the Poincare supergroup. The supergeneralization of the Minkowski metric is the Minkowski supermetric $SM(4,4\vert \lambda, \mu)$ (\ref{SM})  depending on two real parameters $\lambda$, $\mu$ obeying the only condition of nondegeneracy of the supermetric $\lambda\neq \pm \mu$.
    Using the  V.~P. Akulov and D.~V. Volkov's formulae of super-Riemannian geometry  \cite{AkVolk} we  calculated a superconnection and a supercurvature of the Minkowski superspace. It was shown that the curvature of the Minkowski superspace does not vanish,  and the  Minkowski supermetric is the solution of the Einstein superequations (\ref{SEE}),
so the eight-dimensional curved  super-Poincare invariant superuniverse $SM(4,4\vert \lambda, \mu)$  is supported
by purely fermionic stress-energy supertensor (\ref{SET}) with two real parameters $\lambda$, $\mu$, and, moreover, it has non-vanishing cosmological constant $\Lambda=12/(\lambda^2 -\mu^2)$ defined by these parameters that could mean a new look at the cosmological constant problem.

\end{document}